# Core-exsolved SiO$_2$ dispersal in the Earth's mantle

George Helffrich[1,*], Maxim D. Ballmer[2,1], Kei Hirose[1,3]

Abstract. SiO$_2$ may have been expelled from the core directly following core formation in the early stages of Earth's accretion and onwards through the present day. On account of SiO$_2$'s low density with respect to both the core and the lowermost mantle, we examine the process of SiO$_2$ accumulation at the core-mantle boundary (CMB) and its incorporation into the mantle by buoyant rise. Today, if SiO$_2$ is 100-10000 times more viscous than lower mantle material, the dimensions of SiO$_2$ diapirs formed by the viscous Rayleigh-Taylor instability at the CMB would cause them to be swept into the mantle as inclusions of 100 m - 10 km diameter. Under early Earth conditions of rapid heat loss after core formation, SiO$_2$ diapirs of ~1 km diameter could have risen independently of mantle flow to their level of neutral buoyancy in the mantle, trapping them there due to a combination of intrinsically high viscosity and neutral buoyancy. We examine the SiO$_2$ yield by assuming Si+O saturation at the conditions found at the base of a magma ocean and find that for a range of conditions, dispersed bodies could reach as high as 8.5 vol.% in parts of the lower mantle. At such low concentration, their effect on aggregate seismic wavespeeds is within observational seismology uncertainty. However, their presence can account for small-scale scattering in the lower mantle due to the bodies' large velocity contrast. We conclude that the shallow lower mantle (700-1500 km depth) could harbor SiO$_2$ released in early Earth times.

## Introduction

The Earth formed via an accretion process that mixed approximately chondritic materials together and differentiated them into a metallic core and a silicate crust and mantle. In the course of its development, the Earth could have been substantially or entirely melted by a variety of processes: radiogenic heating (Hevey and Sanders 2006), gravitational segregation (Monteux et al. 2009), heating by small impacts (Kaula 1979) or the Moon-forming giant impact (Canup and Asphaug 2001). Melting is the main way a solid body's composition homogenizes (Hofmann and Hart 1978), so the likelihood of past melting events in the Earth's history implies a corresponding likelihood of internal homogeneity within its constituent parts: crust, mantle and core. But the crust and mantle are plainly heterogeneous, as many geologists (and even geochemists and geophysicists) will attest (Allègre and Turcotte 1986). Homogeneity is, nevertheless, a powerful way to characterize the properties of a complex system and forms the first-order view of the Earth's internal structure. The deviations from uniformity — the second-order features — then become the primary source of information from which the details of the evolutionary path may be inferred.

[1]Earth-Life Science Institute, Tokyo Institute of Technology, 2-12-1 Ookayama I7E-312, Meguro-ku, Tokyo 152-8550, Japan; [2]Institut für Geophysik, ETH Zürich, 8092 Zürich, Switzerland; [3]Earth and Planetary Sciences Dept., University of Tokyo, 7-3-1 Hongo, Bunkyo-ku, Tokyo 113-0033 Japan

One source of information is the seismic structure of the mantle. It is broadly peridotitic in composition, one approximation to which is pyrolite (Ringwood 1975), which explains the melting relations of mid-ocean ridge basalt. However, the calculated wavespeeds arising from an equilibrium assemblage of minerals comprising a pyrolitic bulk composition do not explain the mantle's seismic wavespeeds very well (Xu et al. 2008), instead favoring an inhomogeneous mechanical mixture of a basaltic and a harzburgitic composition.

In addition to the theoretical evidence for heterogeneity in the mantle, observations also support the inference. The well-known discrepancy between shear wave travel times calculated from low frequency (normal mode) seismic data and higher frequency body wave data (Dziewonski and Anderson 1981; Nolet and Moser 1993) caused the later authors to propose that small-scale structure at wavelengths shorter than 200 km dispersed in the upper and lower mantle, but with stronger heterogeneity in the upper mantle, could explain it. Gudmundsson et al. (1990) also found evidence in body wave travel times for a less intense heterogeneity in the lower mantle, as did Masters et al. (2000) for 3D tomographic heterogeneity. Tibuleac et al. (2003) reported that the 20-40% variation in amplitudes of broadband P waves recorded across a seismic array required focusing and defocusing of the wavefield by mantle velocity anomalies in excess of 1%, far larger than those of the then extant tomographic models.

At scale lengths much smaller than the 100s of kilometers described above, Hedlin et al. (1997) found evidence for heterogenity distributed in the lower mantle from randomly distributed 1% heterogenties of ~8 km size present throughout the mantle. Subsequent investigation narrowed the extent of the heterogeneity to the lowermost lower mantle extending 1000 km upwards from the core-mantle boundary (CMB) (Hedlin and Shearer 2000), and reducing the level of heterogeneity required by an order of magnitude (Margerin and Nolet 2003; Mancinelli and Shearer 2013). Braña and Helffrich (2004) also found a restricted region near the CMB (a cube of side 700 km) where the small-scale heterogeneity (600 m radius) is particularly intense, suggesting a local source.

In contrast to the statistically distributed heterogeneities described in these studies, Kaneshima and Helffrich (1998; 1999) found small-scale heterogeneities that deterministically scatter seismic waves from nearby earthquakes. These objects are about the thickness of subducted oceanic crust and seem to be organized in planar geometries suggestive of subducted lithospheric fragments. Speculation on the source of the strong velocity anomalies (>8%) required to explain the intensity of the scattered waves led to the idea that the reduction in the shear modulus associated with a second-order phase transition in $SiO_2$ (stishovite to $CaCl_2$ structure) (Carpenter et al. 2000; Bina 2010; Asahara et al. 2013; Xu et al. 2017) may lower the shear modulus in a wide depth range above and below the phase transition pressure. Subsequently, a larger-scale search of scattering intensity around circum-Pacific subduction zones (Kaneshima and Helffrich 2009) showed that the inferred scatterer distribution varied with region and depth, peaking between 1200-1500 km, a range that includes the pressure where the phase transition would occur. However, there appears to be a uniform drop in scattering intensity deeper than around 1600 km despite the method used being sensitive to structure at those depths. This implies that there are fewer scatterers in the lowermost 1200 km of the lower mantle relative to shallower levels.

On account of the implication that $SiO_2$ might be the generative agent for the heterogeneity's seismic visibility, we consider how it might be injected into the mantle. Subduction is an obvious way: various workers showed that mid-ocean ridge basalts can contain up to 10-20 vol% free silica (Ono et al. 2001; Hirose et al. 2005). Deterministic scatterers strong enough for waveform analysis suggest that the scattered waves emanate from objects with the dimensions of subducted crust (Kaneshima and Helffrich 1999; 2003). Other methods for observing scattering found strong sources both close to sites of present subduction and far from it, with some in common to the two methods (Bentham and Rost 2014). Hence whether deterministic scatterers are observable is related to their being close to seismic sources, reflecting a particular methodological bias. In contrast, the background scattering potential seems not to have a similar association with subduction. The scattering potential decays with depth, even though the mantle's depth is equally well illuminated by subduction zone earthquakes as are the deterministic scatterers (Kaneshima and Helffrich 2009). Consequently, we can distinguish at least two distinct scattering sources in the mantle: one related to subduction remnants, and another to an as-yet unidentified process.

This process might have been the exsolution of $SiO_2$ from the core and its expulsion into the mantle. Recent experiments show significant Si solubility in liquid metal (Hirose et al. 2017) at the pressures and temperatures associated with accretion at the base of a magma ocean (O'Neill et al. 1998; Wood et al. 2006) and in the core. As the early core cools, it becomes supersaturated in Si, which takes up dissolved O in the metal and exsolves it as $SiO_2$ (Hirose et al. 2017). Due to its low density compared to liquid metal (Hirose et al. 2017), $SiO_2$ would accumulate at the CMB, which presents a buoyancy discontinuity and a rheological barrier. $SiO_2$'s density is actually lower than that of the seismologically-determined present-day mantle (Hirose et al. 2005), so after a pause for the growth of a diapiric instability, it will further ascend through the mantle until it reaches neutral buoyancy at around 1500-1600 km depth (Wang et al. 2012). Along with convective stirring in the mantle, this could establish a radial profile for heterogeneity in the mantle that could explain the lower mantle's seismic heterogeneity.

In this study, we therefore use geophysical fluid dynamical theory of viscous instability in boundary layers to investigate the dispersal in the mantle of $SiO_2$ released from the core. We employ an equation of state (EOS) for $SiO_2$ to obtain the buoyancy forces that drive the instability and the subsequent incorporation of $SiO_2$ into the mantle. We also consider how the combination of relative density and viscosity contrast serves to isolate $SiO_2$ to the lower mantle, and show that dispersed $SiO_2$ has a minimal effect on aggregate wavespeeds but presents enough of an elastic property contrast with respect to ambient mantle that it scatters seismic waves. The lower mantle's observed small-scale scattering profiles suggest SiO$_2$ aggregation and isolation in the lower mantle.

## Methods

<u>Layer thickness</u>. SiO$_2$ exsolving from the core would accumulate at the CMB in a layer due to the rheological contrast with the solid silicate. The layer thickness is governed by the time that it takes for the layer to become buoyantly unstable. To determine the time scale for the rate of growth of a Rayleigh-Taylor instability at the CMB, we use a modified form of the theory of Ballmer et al. (2017). In it, the critical layer thickness, $b_{crit}$ is given by

$$b_{crit} = \tau_{RT} v_g \quad , \tag{1}$$

where $v_g$ is the rate at which a layer of thickness $b$ at the CMB grows by addition of material expelled by the core (Figure 1), and $\tau_{RT}$ is a characteristic time scale for the onset of the Rayleigh-Taylor instability. $\tau_{RT}$ is generally some function of the medium's material parameters and the boundary geometry divided by the critical layer thickness, $b_{crit}$, that we represent as $f(\cdot)/b_{crit}$. Hence

$$b_{crit}^2 = f(\cdot) v_g \quad . \tag{2}$$

Leaving aside for a moment the specific form of $f(\cdot)$, we focus first on $v_g$. If the concentration by mass of a substance in the core is denoted by $c$, its total mass $M$ in the core is

$$M = M_c \times c \quad , \tag{3}$$

where $M_c$ is the core's mass. If it is expelled at a rate $dM/dt$, then

$$\frac{dM}{dt} = M_c \frac{dc}{dt} = \frac{4}{3}\pi r^3 \bar{\rho} \frac{dc}{dt} \tag{4}$$

if the core has radius $r$ and mean density $\bar{\rho}$. From the definition of the density $\rho$, the change of mass $dM$ is

$$dM = \rho dV = \rho 4\pi r^2 db \tag{5}$$

for a spherical body like the core. If the mass accumulates at the core's surface, the layer thickens at a rate

$$v_g = \frac{db}{dt} = \frac{dM}{dt}\frac{1}{4\pi r^2 \rho} = \frac{\frac{4}{3}\pi r^3 \bar{\rho}\frac{dc}{dt}}{4\pi r^2 \rho} = \frac{r}{3}\frac{\bar{\rho}}{\rho}\frac{dc}{dt} \quad . \tag{6}$$

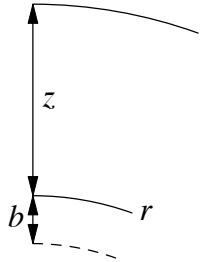

Figure 1. Sketch of mathematical model geometry. Growing layer below CMB (radius $r$) has thickness $b$ ($b \ll r$) and lies below a denser layer of thickness $z$. The material comprising the less dense layer ascends diapirically into the upper layer. The growth time for the diapiric Rayleigh-Taylor instability, $\tau_{RT}$, controls $b$.

On account of $c$'s dependence on temperature $T$ rather than time $t$, $v_g$ may be expressed in terms of the core's cooling rate $dT/dt$:

$$v_g = \frac{r}{3}\frac{\bar{\rho}}{\rho}\frac{dc}{dT}\frac{dT}{dt} \quad . \tag{7}$$

In their earlier analysis, Ballmer et al. (2017) used a 2D Cartesian geometry (Turcotte and Schubert 2002), but here we use a 3D treatment to account for the two-dimensionality of the CMB. Ribe (1998) developed a theory of the Rayleigh-Taylor instability that describes the rate of growth of diapirs forming on a 2D surface when denser material overlies less dense material. In this case,

$$\tau_{RT} = \alpha(k, z/b - 1, \gamma)\frac{\mu_m}{\gamma}\frac{1}{(\rho_m - \rho)gb} \quad , \tag{8}$$

where $\alpha(k, z/b - 1, \gamma)$ is the dimensionless growth rate for dimensionless horizontal wavenumber $k$, $z/b - 1$ is the medium's scaled excess depth (the medium depth $z$ scaled by the layer thickness $b$), $\gamma$ is the ratio of the viscosity of the overlying mantle to the viscosity of the layer forming under it, $\mu_m$ is the viscosity of the overlying mantle and $\rho_m$ is its density, and $g$ is the gravitational acceleration at radius $r$. See Ribe (1998) for the definition of $\alpha(\cdot)$ (eqs. A2 & A3). $k$ is chosen by an optimization scheme that maximizes the growth rate given the geometric and physical factors. Combining (7) and (8), we have, then, an implicit equation for $b_{crit}$,

$$b_{crit}^2 = \alpha(k, z/b_{crit} - 1, \gamma)\frac{\mu_m}{\gamma}\frac{1}{(\rho_m - \rho)g}\frac{r}{3}\frac{\bar{\rho}}{\rho}\frac{dc}{dT}\frac{dT}{dt} \quad . \tag{9}$$

which is solved iteratively initially assuming $b_{crit} = 0$ and calculating successive $b_{crit}$ values until they converge self-consistently.

An alternative mechanism may govern the rate of release of viscous material from a boundary layer. The Rayleigh-Bénard instability (RB) develops when the material in a thermal boundary layer is heated from below and expands until it can buoyantly rise into the overlying convective layer. This instability's time scale is related to thermal diffusion. Following Turcotte and Schubert (2002), the combination of the critical Rayleigh number $Ra_{cr}$ for this case (657.5 for a free surface and 867.8 for free slip) and the critical thermal boundary layer thickness as described by its similarity variable value $\xi_{cr} = \text{erfc}^{-1}(0.01)$ leads to a time scale

$$\tau_{RB} = \left[ \frac{Ra_{cr}\mu\kappa}{g(\rho_m - \rho + \rho\alpha\Delta T)} \right]^{2/3} (4\kappa\xi_{cr}^2)^{-1} \quad . \tag{10}$$

Comparing $\tau_{RT}$ with $\tau_{RB}$, it appears that $\tau_{RB}$ is 10-1000 times longer than $\tau_{RT}$ for any plausible layer growth rate $dT/dt$, viscosity ratio $\gamma$, or intrinsic buoyancy difference $(\rho_m - \rho)$ in present or early Earth times. Hence the RT instability will develop faster than the RB instability, so we may ignore this effect. Thermal instability may contribute to drive and advance the RT instability, but we ignore this effect due to the large difference between $\tau_{RT}$ and $\tau_{RB}$ for any reasonable gamma.

Diapir ascent rate. The diameter of the diapir is taken to be the thickness of the $SiO_2$ layer when it detaches (Ballmer et al. 2017). Bina (2010) showed that due to the reaction of $SiO_2$ with (Mg,Fe)O ferropericlase in the lower mantle, the diapir will become armored with $(Mg,Fe)SiO_3$ bridgmanite that will prevent further reaction. The armoring layer will be 5-10 cm, so we neglect it and the small density change that it causes. We assume that the detached diapir adopts a spherical shape of diameter $b_{crit}$. The effect of entrainment of some ambient mantle with the diapir is small (less than a 15% change in ascent rates for 50% entrained mass) and neglected. The spherical shape assumption is justified experimentally through observed shapes of viscous drops buoyantly moving through a viscous medium. The diapir's position in a Reynolds number-Eötvös number regime diagram indicates it will adopt a spherical shape (Ohta et al. 2010), and initially perturbed shapes with low capillary numbers tend to evolve to sphericity with time (Koh and Leal 1989). (See the supplementary material for the method for estimating surface tension involved with these dimensionless groups.) The density difference relative to the mantle $\Delta\rho = \rho - \rho_m$ will provide the driving force for the rising speed $v$ under Stokes Law (Turcotte and Schubert 2002):

$$v = \frac{2}{9}\Delta\rho g \frac{r^2}{\mu_m} \quad . \tag{11}$$

There is an implicit assumption of a significant viscosity contrast in (11), but the condition is satisfied given the expected material properties (Table 1). Because in (9) $r \sim b_{crit} \sim \mu^{1/2}$ and the ratio $r^2/\mu_m$ appears in (11), $v$ is independent of $\mu_m$ and depends on the viscosity ratio $\gamma$ rather than any absolute viscosity.

The stishovite phase transformation. The stishovite-$CaCl_2$ structured $SiO_2$ is a second-order polymorphic phase transformation that lacks a discontinuity in elastic properties. There is, however, a discontinuity in the pressure and temperature gradient when the transition is crossed. Carpenter et al. (2000) modeled the stishovite-$CaCl_2$ structured $SiO_2$ phase transformation using a Landau model and showed that it affects the shear modulus only. They worked with the individual components of the stiffness tensor $C_{ij}$ and assumed a linear dependence on pressure. Moreover, they analyzed the transformation under isothermal, room temperature conditions. In view of these limitations, here we outline a treatment for the transformation that is both polythermal and amenable to a finite-strain approach.

Table 1. Thermophysical properties for mantle and SiO$_2$

| Property | Symbol | Value | Scale and Units | Source |
|---|---|---|---|---|
| SiO$_2$ yield | $dc/dT$ | 4.1 | $\times 10^{-5}$ K$^{-1}$ | Hirose et al. (2017) |
| Core cooling rate | $dT/dt$ | 100 | K Gyr$^{-1}$ | Nominal rate from Hirose et al. (2017) |
| Thermal diffusivity | $\kappa$ | 1 | $\times 10^{-6}$ m$^2$ s$^{-1}$ | Turcotte and Schubert (2002) |
| Mantle viscosity | $\mu_m$ | 1 | $\times 10^{22}$ Pa s (max) | Lau et al. (2016) |
| | | 1 | $\times 10^{17}$ Pa s (min) | Zimmerman and Kohlstedt (2004) |
| Mantle density | $\rho_m$ | 5560 | kg m$^{-3}$ | PREM$^c$ |
| SiO$_2$ density | $\rho$ | 5460 | kg m$^{-3}$ (CMB) | SiO$_2$ equation of state |
| | $\hat{\rho}$ | 4985 | kg m$^{-3}$ (mean) | |
| SiO$_2$ viscosity ratio | $\gamma$ | 1 | $\times 10^{-4}$ (min) | Xu et al. (2017) |
| | | 1 | $\times 10^{0}$ (max) | $= \mu_m$ |
| Mean core density | $\bar{\rho}$ | 10987 | kg m$^{-3}$ | PREM$^c$ |
| CMB radius | $r_{CMB}$ | 3480 | km | PREM$^c$ |
| CMB temperature | $T_{CMB}$ | 3800 | K | Nomura et al. (2014) upper bound |
| Gravitational acceleration | $g$ | 10.68 | m s$^{-2}$ | PREM$^c$ |
| SiO$_2$ equation of state parameters$^a$ | | | | |
| Reference temperature | $T_r$ | 298.15 | K | |
| Reference volume | $V_0$ | 14.014 | cc mol$^{-1}$ | |
| Thermal expansion coefficients$^b$ | $\alpha_1$ | 2.5 | $\times 10^{-3}$ K$^{-1}$ | Hirose et al. (2017) |
| | $\alpha_5$ | -2.5 | $\times 10^{-2}$ K$^{-1/2}$ | |
| Bulk modulus | $K_0$ | 316 | GPa | |
| Bulk modulus pressure derivative | $K'$ | 4 | | |
| Anderson-Grüneisen parameter | $\delta_T$ | | | $= K'$ (Helffrich and Connolly 2009) |
| Shear modulus | $G$ | 220 | GPa | Stixrude and Lithgow-Bertelloni (2005) |
| Shear modulus pressure derivative | $G'$ | 1.8 | | |
| Phase transition ref. pressure | $P_0$ | 50.2 | GPa | Nomura et al. (2010) |
| Phase transition ref. temperature | $T_0$ | 300 | K | |
| Clapeyron slope | $s$ | 11.1 | $\times 10^{-3}$ GPa K$^{-1}$ | |
| Shear modulus softening | $A_0$ | -145.553 | GPa | Fit to Carpenter et al. (2000). |
| Phase transition width | $w$ | 15.93 | GPa | |

Nomura et al.'s (2010) experiments provide the pressure-temperature dependence of the reaction. We parameterize the transition pressure, $P_{tr}$ as being linearly dependent on temperature:

$$P_{tr}(T) = P_0 + s \times (T - T_0) \quad , \tag{12}$$

with parameters listed in Table 1. The shape of the shear modulus decrease is modeled with a simple functional form that reproduces the cusped transition shown in Carpenter et al. (2000) quite well:

$$\Delta G(P,T) = A_0 \left[ 1 - \frac{2}{\pi} \left| \tan^{-1}([P - P_{tr}(T)]/w) \right| \right]^2 \quad . \tag{13}$$

$\Delta G$ is added to the finite-strain calculated stishovite shear modulus to model its softening through the transition. See Table 1 for the parameters.

Precipitation of $SiO_2$ from the core. We use the $SiO_2$ saturation model of Hirose et al. (2017) to calculate the Si+O content of the metal equilibrated at the bottom of a magma ocean during accretion. Various studies of partitioning of moderately siderophile elements at high pressures defined an effective equilibration pressure which, coupled with an equation for the peridotite solidus, gives the equilibration $P$ and $T$ that explains the present concentration in the mantle and of the core. To allow for a range of models, we use two peridotite solidus equations from Wade and Wood (2005) and from Fiquet et al. (2010). Different experimental studies report a range of effective equilibration pressures (Wade and Wood 2005; Wood et al. 2006; Siebert et al. 2013; Fischer et al. 2015) so we take their range, 30-55 GPa, as bounds on the Si+O saturation level. We then calculate how much $SiO_2$ would be precipitated from the core to saturate it at the present CMB conditions, 3800 K and 135 GPa (Table 1). Depending on the initial Si+O content, a range of $SiO_2$ yields is possible even at the same magma ocean $P$ and $T$ conditions. Hence we track the low, high and mean values of $SiO_2$ yield.

The mass of $SiO_2$ expelled by the core is then turned into a volume fraction in the lower mantle. We use a linear increase in volume fraction with height above the CMB that reaches a plateau at $r_m$, above which it stays constant, or a peak at $r_m$ after which it returns to zero at $r_l$. (The reasons for this choice will become clear in the Discussion section.) Hence the volume fraction $f$ is

$$\begin{aligned} f(r) &= f_{max} \times \min(1, (r - r_{CMB})/(r_m - r_{CMB})) && \text{(plateau)} \\ f(r) &= f_{max} \times \begin{cases} (r - r_{CMB})/(r_m - r_{CMB}) & , r < r_m \\ \max(0, 1 - (r - r_m)/(r_l - r_m)) & , r \geq r_m \end{cases} && \text{(peak)} \end{aligned} \tag{14}$$

If the mass fraction of $SiO_2$ crystallized is $\Delta m_{SiO_2}$, then the mass of $SiO_2$ is $M_{core} \times \Delta m_{SiO_2}$. Dispersing this in the lower mantle, the volume $V_{SiO_2} = M_{core} \Delta m_{SiO_2}/\hat{\rho}$

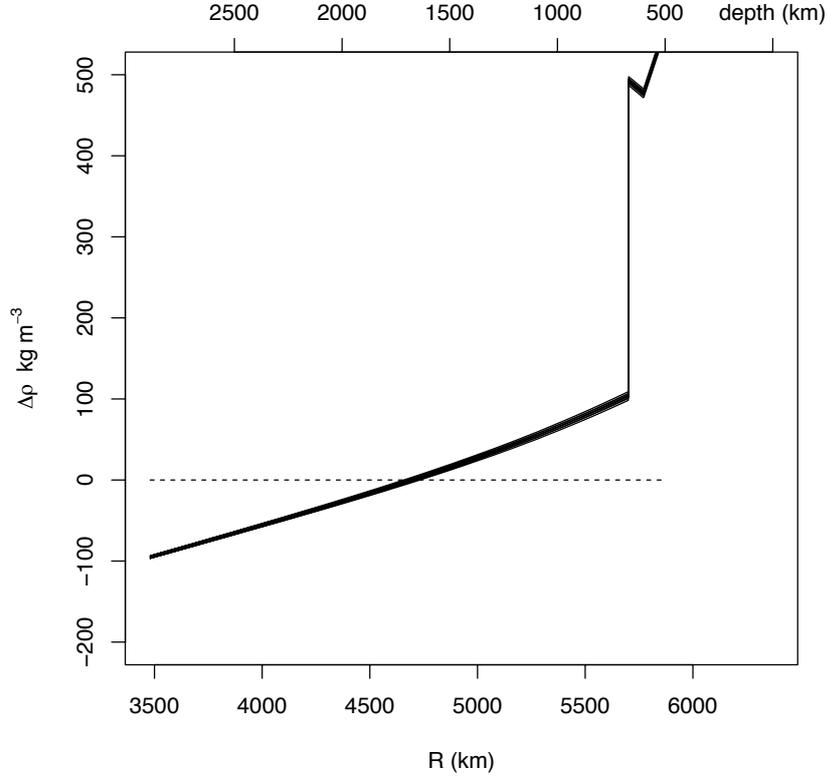

Figure 2. $SiO_2$ density difference with respect to the PREM in the lower mantle. $\Delta\rho = \rho - \rho_m$ calculated using $SiO_2$ equation of state information (Table 1) along various adiabats initiated above the CMB. CMB temperature is assumed to be 3800 K with a 1050±50 K thermal boundary layer above it to project adiabatic temperatures to ~1950 K at 660 km depth (Katsura et al. 2010) using a variety of lower mantle Grüneisen parameters $\gamma$ with $1.3 \le \gamma \le 1.5$ (Helffrich 2017). $SiO_2$ becomes neutrally buoyant at a depth of ~1600 km.

and so the lower mantle volume fraction is

$$f_{SiO_2} = V_{SiO_2}/V_{LM} = \frac{M_{core} \times \Delta m_{SiO_2}}{\hat{\rho} V_{LM}} \quad , \tag{15}$$

where $\hat{\rho}$ is the mean density of $SiO_2$ along the adiabats calculated in the lower mantle. We want the $f_{max}$ from (14) that yields the calculated $f_{SiO_2}$ which is the volume weighted integral of $f(r)$ through the lower mantle. Sparing the reader some algebra, the scaling factors are numerically

$$f_{\max} = f_{SiO_2} \times \begin{cases} \dfrac{1}{0.773} & \text{(plateau)} \\ \dfrac{1}{0.515} & \text{(peak)} \end{cases}. \qquad (16)$$

## Results

The EOS for $SiO_2$ (Table 1) provides its density in the lower mantle, and may be compared with PREM's density (Dziewonski and Anderson 1981). At present-day CMB conditions, the contrast is about 100 kg m$^{-3}$ (~2%) less dense than the mantle. This is the origin of the buoyancy force that gives rise to the diapiric instability. Figure 2 shows the density contrast with respect to PREM's mantle, which lessens with height above the CMB. At depths shallower than 1600 km, $SiO_2$ becomes denser than ambient mantle and the buoyancy force changes sign. In the upper mantle, the density contrast increases by a factor of five, enhancing the tendency of $SiO_2$ to sink.

Using the critical layer thickness calculated with equation (9) one can examine its dependence upon mantle and layer viscosity. $SiO_2$'s viscosity may be bracketed on one hand by assuming that it is the same as the lower mantle's viscosity, $10^{22}$ Pa s (Table 1), and on the other hand through viscosity-diffusivity systematics applied to stishovite (Xu et al. 2017; Jaoul 1990) that indicate it is possibly $10^4$ times more viscous (Table 1). Large viscosity ratios lead to larger $SiO_2$ bodies ascending from the CMB through the mantle, whereas ratios closer to 1 yield meter-to-kilometer sized $SiO_2$ bodies. Figure 3 shows this dependence for present day Earth-like values (Table 1). Depending on the viscosity ratio values, the critical layer thickness may range from 80 m to up to 30 km, though 1-30 km is more probable. These would be unable to rise through the convective mantle. Using Dumoulin et al's (2005) parameterization to estimate viscosities in early Earth conditions, we conservatively take a mantle temperature 500 K hotter than present (see Lebrun et al. (2013) for a variety of estimates), which leads to viscosities $\geq 10^3$ lower than now and 10 m - 2 km size diapirs.

A comparison with present-day mantle convection is warranted because according to conventional scaling laws, convection speeds scale as $Ra^{2/3}$ (Turcotte and Schubert 2002), where $Ra$ is the thermal convection Rayleigh number. Since $Ra \sim \mu^{-1}$, mantle convection speeds scale as $\mu^{-2/3}$. In contrast, the Stokes speed scales as $\mu^{-1}$, so even in early Earth conditions when the mantle was hotter (higher Rayleigh number) and less viscous, the Stokes speed increases faster with decreasing $\mu$ than convection speeds. Characteristic convection speeds in the mantle are ~1.5 cm yr$^{-1}$ at present (Becker et al. 1999), and maximum rates correspond to the descent rates of subducted lithospheric slabs, ~5 cm yr$^{-1}$ (Bina 2010). For bodies of $SiO_2$ to buoyantly rise through and segregate in the mantle, a combination of entrainment factors and diapir sizes to yield $v$ in excess of the convection speed is needed. Figure 4 examines this tradeoff. It depicts the viable combinations given the present conditions at the Earth's CMB (Table 1) in terms of $\gamma$. Given present-day cooling rates of 50-100 K/Gyr (Herzberg et al. 2010; Hirose et al. 2017) and expected viscosity ratios, rising speeds are 10-1000 times slower than mantle convection speeds. Hence the bodies would be drawn passively through the mantle,

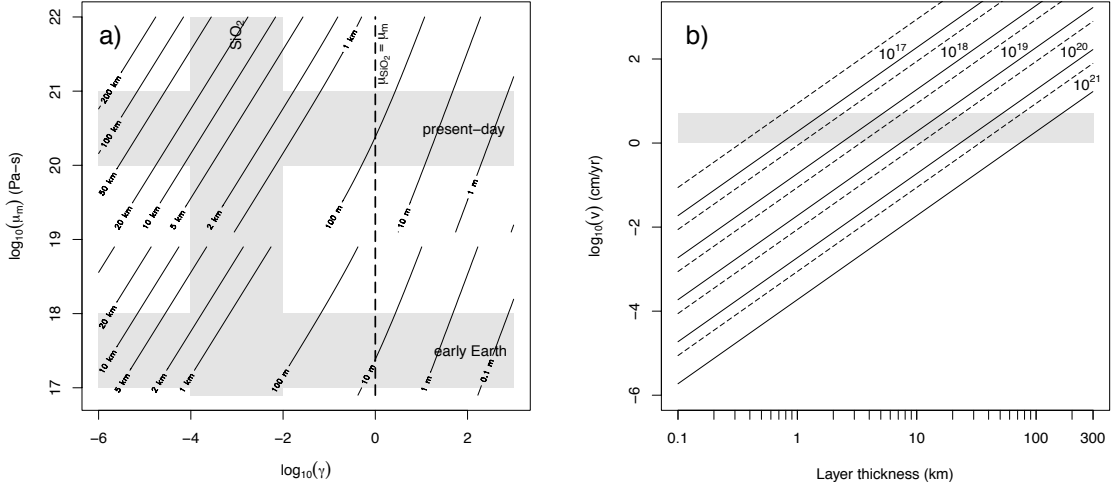

Figure 3. (a) Critical SiO$_2$ layer thickness for present and early Earth CMB conditions. $\gamma$ is the ratio $\mu_m / \mu$ of the mantle's viscosity to the SiO$_2$ layer's. Vertical shaded band brackets probable range of SiO$_2$ viscosity ratio, while horizontal bands indicate likely $\mu_m$ range. Layer thickness might grow to as much as 0.08-30 km at present-day lower mantle viscosities and 100 K/Gyr cooling rate if the layer is 1-10$^4$ times more viscous ($-4 \leq \log_{10}(\gamma) \leq 0$), and 10 m - 2 km in early Earth conditions at 1000 K/Gyr cooling rate. (b) Stokes ascent speed (solid lines) for diapirs detaching from layers of various thicknesses at various $\mu_m$ values (Pa-s). Descent speeds of dense SiO$_2$ bodies in the upper mantle (dashed lines) shown for same ranges of $\mu_m$. Shaded band shows maximum vertical convection speed estimates. Present-day viscosities require layer thicknesses >20 km to rise independently of convective speeds. In the early Earth epoch when viscosities were lower, bodies >~ 800 m could detach and rise independently, and the high sinking rates for dense bodies as small as 300 m would confine them to the lower mantle. At present, bodies >2-3 km would also be confined to the lower mantle.

initially entrained by the flow in the mantle at the CMB and not ascend through it.

In order to overcome the viscous drag resisting ascent, SiO$_2$ accumulating at the CMB needs to detach either as large bodies caused by rapid cooling rates (upper left in Fig. 3a) or due to reduced mantle viscosities caused by higher mantle T. In the early Earth, heat flows from the core were probably much higher. Simulations of convection in a crystallizing magma ocean suggest cooling rates of 300 K/Gyr when heating is moderated by the development of an atmosphere that thermally blankets the surface (Lebrun et al. 2013). In contrast, simulations of the initial phases of magma ocean development before an atmosphere outgasses (or is removed by continuing impacts during accretion) lead to cooling rates of 2000 K/Gyr (Solomatov and Stevenson 1993). An extreme case is heating of the core by gravitational potential energy liberated by descending dikes or diapirs of metal segregating from a shallower magma ocean or impact-formed magma lake (Stevenson 2003; Monteux et al. 2009), which could leave the growing outer core hottest at the CMB and, potentially, thermally stratified (Lasbleis et al. 2016). The solution to the one dimensional heat equation (Turcotte and Schubert 2002) leads to a cooling rate of

$$\frac{dT}{dt} = \frac{-\Delta T z}{2t\sqrt{\pi \kappa t}} \exp\left[-\frac{z^2}{4\kappa t}\right] \quad , \qquad (17)$$

where $\Delta T$ is the initial temperature difference between the core and the mantle and $z$ is distance from the CMB. For an initial temperature difference of 4000 K and characteristic distance of 1 km from the CMB, cooling rates at 0.1 Gyr, 0.01 Gyr and 0.001 Gyr are 200 K/Gyr, 6300 K/Gyr and 20,000 K/Gyr. We choose a cooling rate >1000 K/Gyr to show the consequences of earlier, more rapid cooling and a higher temperature mantle. Under these conditions, $SiO_2$ bodies are likely to be small, certainly < 10 km in size and more probably ~1 km (Fig. 4b). Though small, the low mantle viscosity permits a sufficiently rapid rise given $SiO_2$'s expected viscosity contrast.

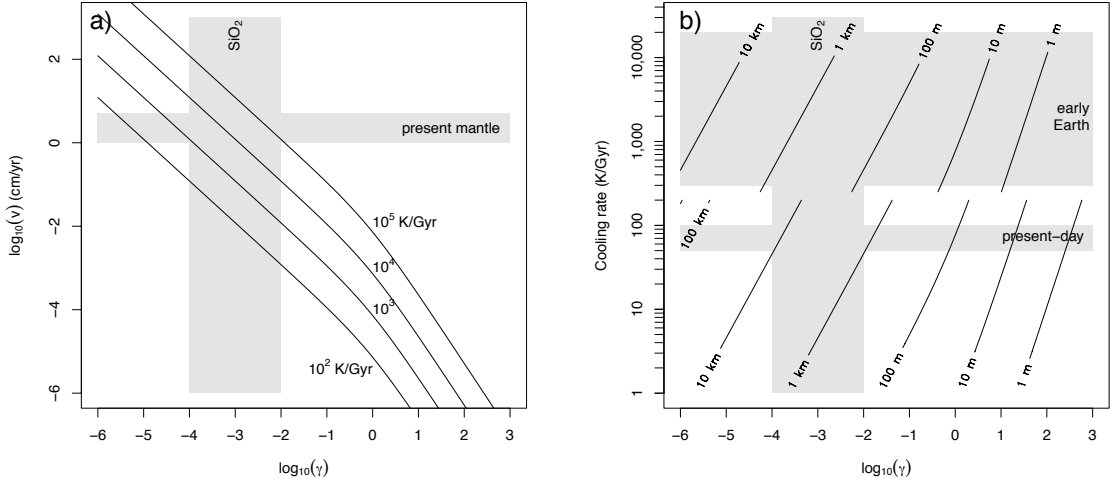

figure 4. (a) Ascent rate of $SiO_2$ in the mantle as a function of the viscosity ratio $\gamma$ for a suite of representative core cooling rates (labels; present-day is ~$10^2$ K/Gyr). In order to exceed the background convective speed of the mantle, 1-5 cm/yr (horizontal gray band), $SiO_2$ viscosities would need to exceed the highest expected contrasts ($> \mu_m \times 10^4$), or rapid cooling due to early Earth conditions would be required. (b) $SiO_2$ diapir diameter for different viscosity ratios $\gamma$ and different cooling rates. Present-day cooling rates lead to bodies unable to rise through background mantle convection, whereas early Earth rates >1,000 K/Gyr lead to independent rise of somewhat smaller bodies. Present-day and early Earth $\mu_m$ used in the calculation are $10^{20.5}$ and $10^{17.5}$ Pa-s, respectively. Vertical gray band in each panel shows likely $SiO_2$ viscosity contrast range.

Due to the large negative density contrast in the upper mantle (Figure 2), the bottom of the transition zone acts as a strong filter against $SiO_2$ entry. Bodies 2-3 km in radius would sink in background convective flows of 1-5 cm/yr (Fig. 3b) if upper mantle viscosity is $10^{19}$ Pa-s, a plausible upper mantle viscosity even now (Table 1). In early Earth times, mantle temperatures 100-200 K warmer than present would reduce viscosities by factors of 100, again using Dumoulin et al.'s (2005) viscosity scaling. The viscosity would probably be significantly lower than this (Fig. 3a), ensuring $SiO_2$'s initial confinement to the lower mantle, and arguably maintaining it to the present.

## Discussion

The scattering potential in five circum-Pacific subduction zones estimated by Kaneshima and Helffrich (2009) may be compared with the $SiO_2$ dispersal characteristics that were derived earlier. They exhibit two prominent features (Figure 5): a drop-off in scattering power below 1500 km depth; and a peak or knee in their profiles at 1500-1800 km depth. Figure 2 shows that $SiO_2$ is neutrally buoyant at ~1600 km. However, Figure 4a shows that $SiO_2$ evolved at the CMB at present would lack sufficient buoyancy to rise independently of ambient mantle flow. Rather, it would be a tracer of flow in the mantle that was once in contact with the CMB.

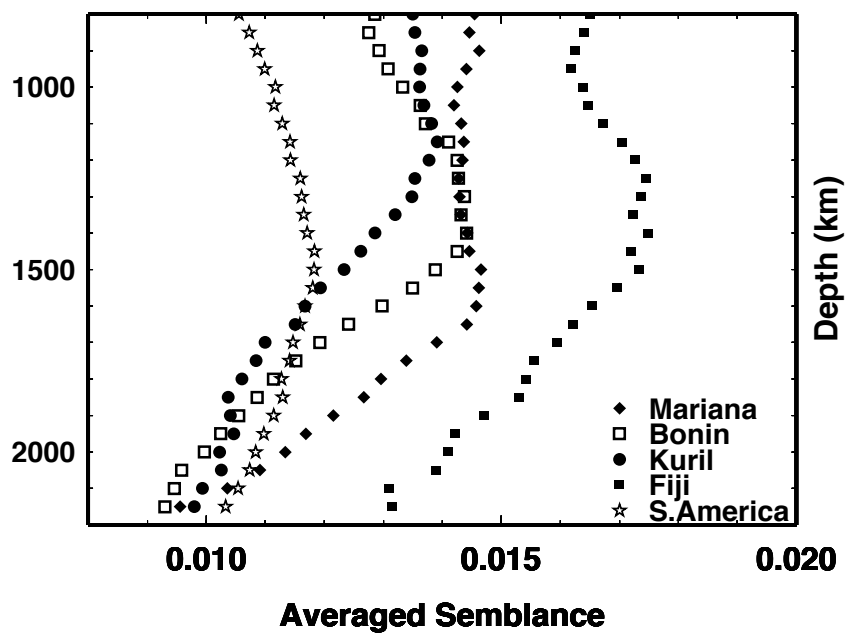

Figure 5. Scattering potential as a function of depth in five circum-Pacific subduction zones. Semblance, the ratio of coherently stacked power to incoherently stacked power, is a proxy for the scatterer distribution. Except for the Kuril subduction zone, the profiles show either a peak or a dropoff of scattering potential at 1500-1700 km depth. See Kaneshima and Helffrich (2009) for details.

Another inference that may be drawn from the profiles is from the scattering power dropoff below 1500-1800 km. If this were due to dispersal of small bodies created at the CMB, their concentration would be highest at the CMB and would decrease upwards. Because the scattering intensity increases with distance from the CMB, it is not likely to be created by this process. Moreover, due to the passive entrainment of any $SiO_2$ emerging through the CMB, the profiles also do not seem to be generated by a process acting at the present day.

The rough coincidence in four of five cases of the knee in the scattering potential curve with the neutral buoyancy level of $SiO_2$ suggests that the bodies must be able to raise themselves to that level. Yet if they were large enough to do this (30-200 km in diameter given present mantle conditions; Fig. 3), they would probably be seismically visible either as tomographic anomalies or as deterministic scatterers. Because deterministic scattering can detect objects as small as 8 km (Kaneshima and Helffrich 1999), we think this unlikely and rather appeal to early Earth conditions when the mantle was partially or wholly molten but $SiO_2$ was not (see melting curves of mantle minerals investigated by Shen and Lazor (1995) for example), which would lead to a large viscosity contrast with respect to the mantle. Then (Fig. 3), even small, kilometer-sized objects could rise rapidly through the mantle due to the large viscosity contrast between a partially molten mantle and a solid, inviscid $SiO_2$ body. The objects must remain at that level through to the present day, however, for this to be a viable mechanism. Their intrinsic density contrast prevents them from entering the upper mantle due to the increase of the density contrast by a factor of five and a decrease of the viscosity by a factor of 10, leading to a factor of 50 increase in settling rate (equation (12) and Fig. 3b). This effectively prevents $SiO_2$ bodies from being homogenized by melting in the mid-ocean ridge basalt source region. Manga (1996; 2010) showed that viscous bodies dispersed in a convecting medium tend to aggregate, forming grouped heterogeneities from smaller ones. The bodies, though neutrally buoyant, do not merge due to their higher viscosity's diversion of flow around them, inhibiting coalescence. Becker et al. (1999) further investigated the way that viscous materials could survive in the mantle for long times. They showed how viscous bodies with an intrinsic density contrast resist stirring and dispersal in the mantle. This is presumably the mechanism by which the $SiO_2$ bodies stay in the lower mantle at present near their neutral density level. Hence we investigate whether scatterer concentrations that peak in the mid-lower mantle are compatible with the lower mantle's seismic profile.

There are two main ways to examine the homogeneity of the lower mantle directly. The obvious way is to calculate the density, bulk ($K$) and shear ($G$) moduli of an aggregate of PREM-like material and $SiO_2$ and compare them to the PREM uncertainties associated with those quantities (Masters and Gubbins 2003). A more subtle way is to apply Birch's homogeneity index (Birch 1952) to the aggregate composite profile. We use both approaches in the following.

To calculate aggregate properties of PREM and $SiO_2$ we calculate $K$ and $G$ from PREM and the $SiO_2$ EOS augmented with the shear modulus softening due to the stishovite to $CaCl_2$-structured $SiO_2$ phase transition (see Methods). The mean of the Hashin-Shtrikman (HS) bounds on the aggregate $K$ and $G$ (Watt et al. 1976) is used to represent the bulk properties. (For these materials the HS bounds are quite close, so the use of their mean is justified.)

Birch's homogeneity index (also called the Bullen parameter $\eta_B$ by Dziewonski and Anderson (1981)) derives from the relationship he found between the pressure derivative of the bulk modulus and squared bulk sound speed $\Phi = K/\rho$. In Dziewonski and Anderson's (1981) notation,

$$\eta_B = \frac{dK}{dP} + \frac{1}{g}\frac{d\Phi}{dr} \quad . \tag{19}$$

$\eta_B \approx 1$ signifies uniform self compression. It is worthy of note that $\eta_B$ does not depend on $G$ and hence is unaffected by the phase transition in $SiO_2$. The aggregate's properties yield values for $K$ and $\Phi$, and $P$ and $g$ are calculated from PREM's radial density profile.

We use a simple parameterization of the scattering profiles shown in Figure 5 that assumes that the scattering potential is linearly related to the volume fraction $f$ of $SiO_2$ in the mantle, and that $f = 0$ at the CMB. The volume fraction increases linearly with radius until 1500 km depth, whence it attains a constant value $f_{max}$ in the rest of the lower mantle (the plateau model) or peaks at $f_{max}$ and returns to zero at the top of the lower mantle (the peak model). Figure 6 presents a sketch of this relation.

The aggregate properties are calculated from PREM and the $SiO_2$ EOS (Table 1) along the adiabat used for the $SiO_2$ density calculation (Figure 2). Figure 6 shows the results of a direct $K$ and $G$ comparison and the application of the homogeneity test. For plateau model values of $f_{max} \leq 0.2$, both the moduli and $\eta_B$ are within PREM's uncertainties. Hence comparison with PREM limits volumetric heterogeneity in the mantle to less than 20%.

Actual volume fractions will be substantially lower than this, however. A limit to the total mass of $SiO_2$ expelled by the core as it cools may be obtained from the saturation level of Si+O in the metal that accumulated at the bottom of the early Earth's magma ocean. Figure 7 shows the maximum volume fraction of $SiO_2$ in the mantle obtained this way, using a variety of initial Si+O compositions, magma ocean depths, and temperatures at the base of the magma ocean (see Methods). The limits are expressed in terms of $f_{SiO_2}$ and $f_{max}$ for the two scatterer distribution models (see eq. (14) and Fig. 6). The potential limits range from a minimum of no $SiO_2$ to a maximum of ~8.5% by volume, with mean amounts between 1 and 3%. Given PREM uncertainties, this quantity could easily be hidden in the lower mantle.

The actual velocity contrast between the $SiO_2$ and the ambient lower mantle material will still be quite strong, however. Figure 8 shows the variations in the material properties density and Lamé parameters that affect scattering (Wu and Aki 1985). Studies of lower mantle scattering report visible signals from material contrasts of 0.1-0.2% (Margerin and Nolet 2003; Mancinelli and Shearer 2013) to 4-8% (Kaneshima and Helffrich 1999). The $\lambda$ and $\mu$ variations are larger than any of these values. The $\delta\lambda$ curve, in particular, closely approximates the shape of the observed scattering potential (Figure 5), and possibly provides a way to characterize the size of the scattering bodies and their actual

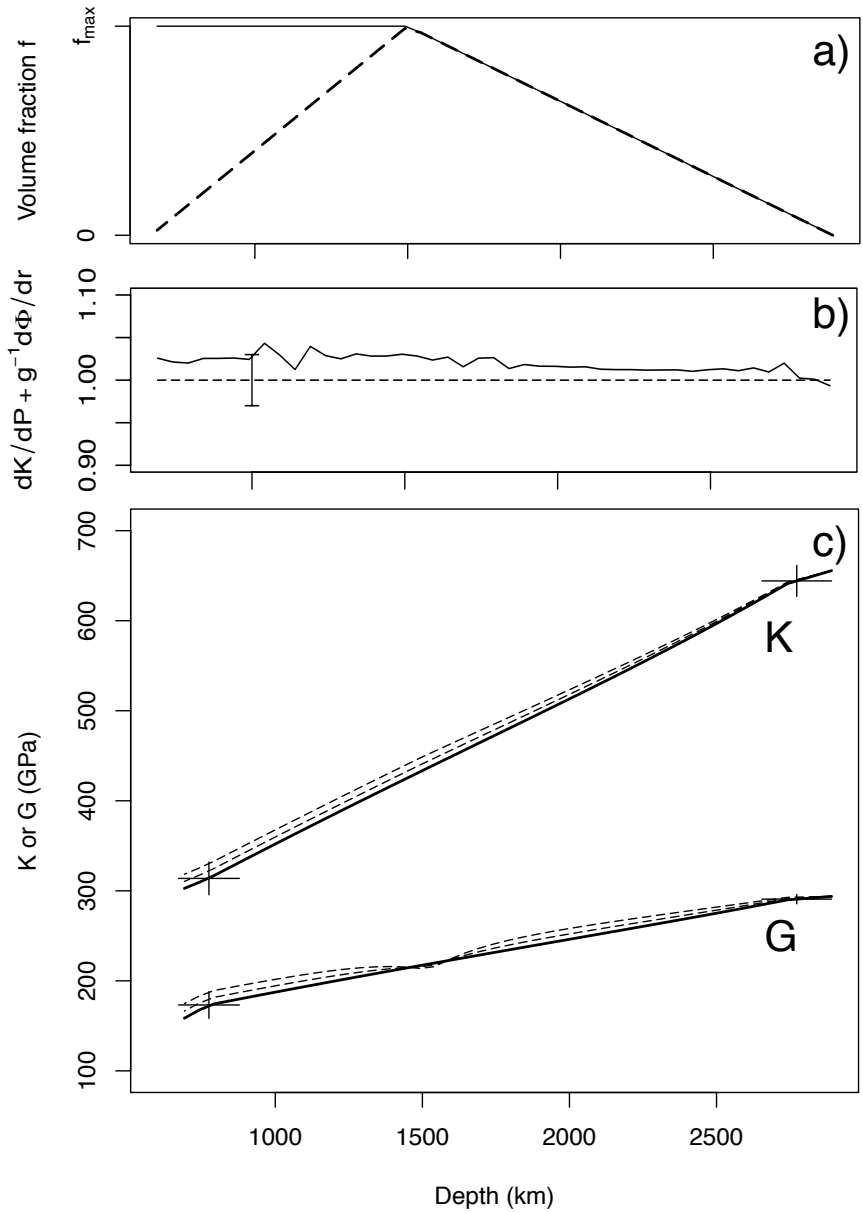

Figure 6. Behavior of a conjectured scatterer density profile through the lower mantle. Top panel (*a*) shows two models of scatterer volume fraction through the lower mantle inspired by the profiles shown in Fig. 5. Volumetric scatterer fraction grows from zero at the CMB to a plateau value, $f_{\max}$, at 1500 km depth (solid), or reaches a maximum of $f_{\max}$ and returns to zero at the top of the lower mantle (dashed). Middle panel (*b*) shows Bullen's homogeneity index (Dziewonski and Anderson 1981) as a function of depth for plateau model when $f_{\max} = 0.2$. A value of 1 indicates uniform self-compression. Error

bar shows PREM variation in homogeneity index in the lower mantle, showing that for volume fractions $f_{max} \leq 0.2$ the index is indistinguishable from PREM; oscillations in trace represent numerical noise in fit. Bottom panel (*c*) bulk (*K*) and shear modulus (*G*) dependence in the lower mantle for PREM (solid lines) and mixtures (dashed lines) of PREM and $SiO_2$ along a lower mantle adiabat for plateau models with $f_{max} = 0.1$ and $0.2$. Depression in $G$ around 1500 km depth is due to $SiO_2$ phase transition. Error bars at tops and bases of profiles are 95% confidence level for the moduli.

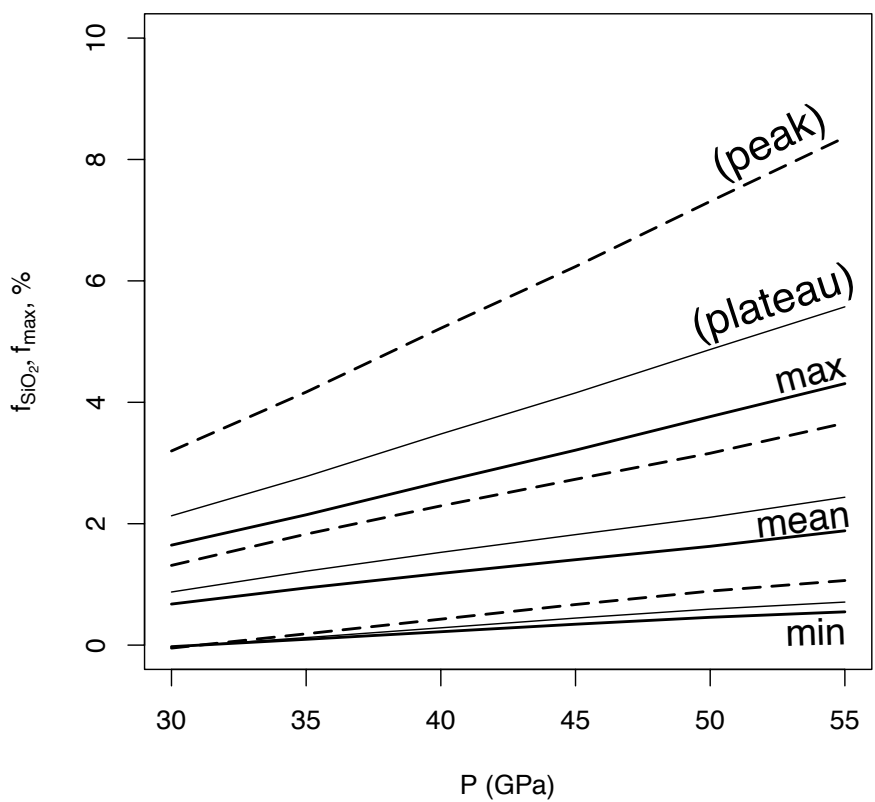

Figure 7. Volume fraction of the mantle taken up by $SiO_2$ expelled from core the between the magma ocean era and the present. Saturation level of Si+O at the base of the magma ocean, set by the pressure at its base, is reduced by the cooling of the core, which crystallizes $SiO_2$ that is incorporated into the mantle. Line trios show minimum, mean, and maximum volume fraction (thick solid) $f_{SiO_2}$ of $SiO_2$ in the lower mantle, and $f_{max}$ for peak (dashed) and plateau (thin solid) models. Depending on the initial Si+O content and scatterer distribution model, the volume fraction may be between essentially zero and ~8.5%.

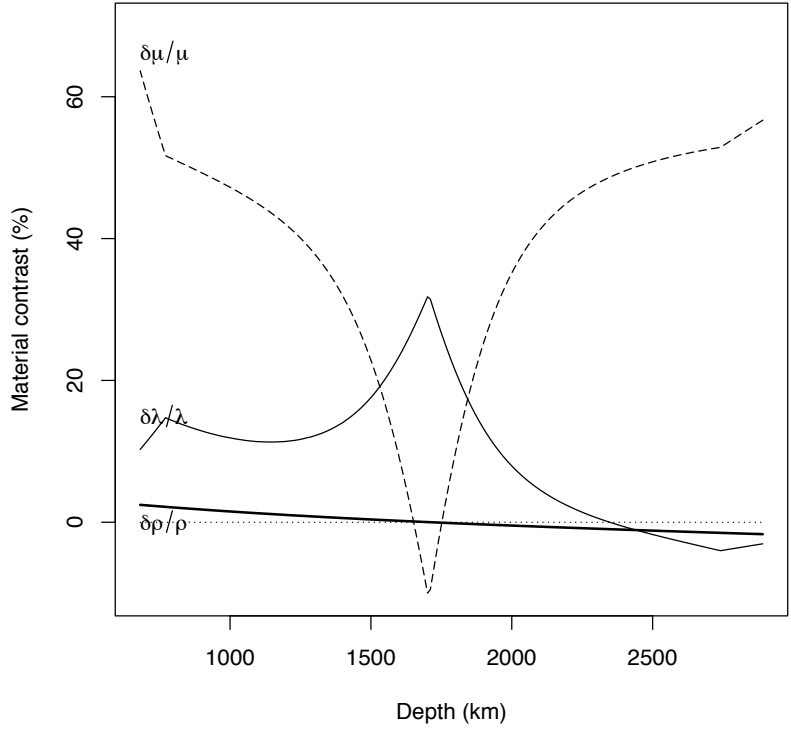

Figure 8. Calculated material property variations between $SiO_2$ and PREM as a function of depth along an adiabat used to calculate the density anomaly (Fig. 2). Scattering is affected by variations in the material properties of density, $\delta\rho/\rho$ and the Lamé parameters $\delta\lambda/\lambda$, $\delta\mu/\mu$. Sharp drop in $\delta\mu/\mu$ and peak in $\delta\lambda/\lambda$ is due to $SiO_2$ phase transition.

material contrasts. This will be investigated in a future study, however.

The increased viscosity of $SiO_2$ relative to bridgmanite and its presence in the lower mantle might also play a role in the well-known viscosity increase in the lower mantle relative to the upper (Haskell 1935). Becker et al. (1999) examined the consequences of a viscosity increase on mantle convection. Using their relation for the effective viscosity $\mu_e$ of a medium composed of more viscous spheres dispersed in a viscoelastic medium of viscosity $\mu_m$,

$$\frac{\mu_e}{\mu_m} = \frac{1}{6}\left[(5f-2)\gamma + (3-5f) + \left[24\gamma + ((5f-3) + (2-5f)\gamma)^2\right]^{1/2}\right] \; , \qquad (20)$$

we can estimate the volumetric abundance $f$ of $SiO_2$ from the increase in viscosity of the lower mantle, which is a factor of 10-100 more viscous than the shallower mantle (Lau et al. 2016). However, volume fractions between 35-40% of material $10^2$-$10^4$ more viscous than the mantle are required to raise the mantle's viscosity by factors of 10-100, much higher than the limits found by either PREM's uncertainty (<20%) or the $SiO_2$ yield capacity of the core (<8.5%). Hence other factors, such as the lower mantle's intrinsic mineralogy of $(Mg,Fe)O$ ferropericlase and $(Mg,Fe)SiO_3$ bridgmanite, must be the main agent for the increased viscosity rather than any excessive concentration of $SiO_2$.

The inferences drawn in the discussion is so far assume that the Rayleigh-Taylor instability sets the governing time scale for $SiO_2$ dispersal. If other instability mechanisms come into play, some revisions may be warranted. We also assume that the higher viscosity of $SiO_2$ allows it to mantain its position near the neutral buoyancy level today, even though it reached that level during early Earth conditions, through the bodies' collective effect on convective flow in the mid-lower mantle. The models for this are only semiquantitative, and when addressed in greater detail, could lead to firmer predictions of the radial dispersal patterns of $SiO_2$ in the lower mantle, and whether $SiO_2$ is able to penetrate the upper mantle density barrier. We also assume that there is sufficient Si+O uptake by the metal en route to the core to both power the dynamo and expel significant amounts of $SiO_2$ during the early Earth era. Geochemical estimates of the core's composition easily satisfy this requirement, however (Hirose et al. 2017), rendering the assumption reasonably secure.

**Conclusions**

We examined the ways by which $SiO_2$, expelled from the core as it cools, could be incorporated into the mantle and be detected today. We developed a model that relates the rate of expulsion of $SiO_2$ due to core cooling to the characteristic scale of a growing viscous Rayleigh-Taylor instability at the CMB. The scale governs the way $SiO_2$ gets dispersed in the mantle. Assuming that $SiO_2$ is 100-10,000 times more viscous than bridgmanite, justified by recent experimental work on their relative volumetric diffusion rates, we find that at rates of present-day core cooling any $SiO_2$ evolved would be in bodies unable to ascend independently of the ambient mantle flow. Hence they would act as passive tracers in the mantle and would not preferentially accumulate due to buoyancy effects. However, early in Earth's formation history when the core was cooling more rapidly, $SiO_2$ would have accumulated in bodies of sufficient size, ~1 km, to move independently of mantle flow and seek their level of neutral buoyancy, which is ~1600 deep in the mid-lower mantle. Once segregated near their neutral buoyancy level, they would resist stirring due to their increased viscosity relative to the surrounding mantle. Whenever they formed, they would be confined to the lower mantle by their density contrast with the upper mantle.

We also determined the limits to the volumetric proportion of $SiO_2$ bodies in the lower mantle given the uncertainties in the whole-Earth model PREM and from the saturation level of Si+O in the core anticipated in magma ocean conditions. The PREM uncertainties limit the volume fraction of $SiO_2$ heterogeneity to values less than 20%, while the Si+O contents to even lower values, ~8.5%. The material property contrast of individual $SiO_2$ bodies with respect to the ambient lower mantle is quite large, however, and the shape of the $\delta\lambda$ depth dependence strongly resembles scattering intensity profiles near circum-Pacific subduction zones and warrants future study to assess whether typical scattering geometries are sensitive to the property variation. The presence of more viscous lower mantle bodies, however, does not appreciably affect the bulk viscosity of the lower mantle, contributing at most a factor of two increase to it.

Acknowledgements. We acknowledge, with thanks, Satoshi Kaneshima, who provided Figure 5, and Matthieu Laneuville for providing useful comments. The reviewers' insights led to some key clarifications to the manuscript, for which we thank them. This work partially funded by JSPS Kakenhi grant JP16H06285 to H.K. and MEXT Kakenhi grant 15H05832 to G.H. Figures and calculations made using R (R Core Team 2017).

# Supplementary Information

<u>Estimate of surface tension for $SiO_2$</u>. Two dimensionless groups that affect the shape of a viscous body moving buoyantly in a viscous medium are the capillary number and the Eötvös number. The capillary number is the ratio of viscous forces to surface tension,

$$Ca = \frac{\mu v}{\sigma} \quad , \tag{S1}$$

while the Eötvös number is the ratio of buoyancy forces to surface tension

$$E\ddot{o} = \frac{\Delta \rho g d^2}{\sigma} \quad . \tag{S2}$$

In both definitions, $\sigma$ is the surface tension, which is problematic to define for a solid. Therefore we use an order-of-magnitude approach, acknowledging that it will be incorrect in some details, but will provide a useful estimate.

We start with the Helmholz free energy $F$, whose total derivative in a system of constant composition is (Hansen and McDonald 2013)

$$dF = -SdT - PdV + \sigma dA \quad , \tag{S3}$$

where $V$ is volume, $P$ is pressure, $S$ is entropy and $T$ is temperature and $A$ is area. Assuming isothermal conditions ($dT = 0$) for simplicity, at equilibrium ($dF = 0$),

$$\sigma = \frac{PdV}{dA} = \frac{\Delta W}{\Delta A} \quad . \tag{S5}$$

This says that the surface tension is the work required to change the system's area.

We can use some simple finite strain relations to develop a relationship between $V$, $P$ and $A$. The finite strain parameter $f$ controls the change in volume of a substance from reference conditions ($P_0, V_0$)

$$f = \frac{1}{2}\left[(V/V_0)^{-2/3} - 1\right] \quad , \tag{S6}$$

and the Murnaghan equation controls the change in volume with a change in pressure

$$V/V_0 = (1 + K'P/K_0)^{-1/K'} = (1 + 2f)^{-3/2} \quad . \tag{S7}$$

Rearranging (S7), an expression for $P(f)$ is

$$P = \frac{K_0}{K'}\left[(1 + 2f)^{3K'/2} - 1\right] \quad . \tag{S8}$$

The use of $f$ simplifies the expression for the variation in pressure and volume throughout a planet; in the Earth's mantle pressure range, $0 \leq f \leq 0.3$ (Helffrich 2017), so volumes shrink by about 50%. (S8) may be integrated to provide the work $\Delta W$:

$$\Delta W = \int_{V_0}^{V} P dV = \int_{0}^{f} P(f)(dV/df) df \qquad (S9)$$

$$= -\frac{V_0 K_0}{K'} \left[ (1+2f)^{-3/2} + \frac{1}{K'-1}(1+2f)^{3(K'-1)/2} \right]_0^f .$$

Given SiO$_2$'s properties (Table 1), to leading order, this expression is $K_0 \times V_0$, which corresponds to $-4.5 \times 10^6$ J. Using the definitions of area and volume for a spherical object, we can express a change in surface area in terms of $f$ as well:

$$A = (6\sqrt{\pi}V_0)^{2/3}(1+2f)^{-1} . \qquad (S10)$$

For an area change to conditions from the surface to the base of the mantle, $\Delta A = -1 \times 10^{-3}$. Hence $\sigma \approx 4.5 \times 10^9$ N m$^{-1}$.

Applying this estimate to $Ca$ (S1), using viscosity estimates from Table 1 and buoyant rise speeds of $10^{-6} \leq v \leq 10^1$ cm/yr (Figure 5), we obtain $Ca$ as low as $10^{-7}$ and as high as 1. For $E\ddot{o}$ (S2), we estimate values (Table 1) of $\approx 1$. From the regime diagram given in Ohta et al. (2010), we can expect spherical shapes due to the low $E\ddot{o}$, and from the low $Ca$ we expect that initially non-spherical shapes will evolve into roughly spherical ones (Koh and Leal 1989), justifying the use of an approximately spherical body.